\documentclass[journal]{IEEEtran}

\ifCLASSINFOpdf
\else
   \usepackage[dvips]{graphicx}
\fi
\usepackage{url}
\usepackage{graphicx}
\usepackage{amsmath,amsfonts,mathtools}
\usepackage{cite}

\begin{document}

\title{Fast algorithms\\ for complex-valued discrete Fourier transform\\ with separate real and imaginary inputs/outputs}

\author{Aleksandr Cariow
\noindent \thanks{Aleksandr Cariow is with the Faculty of Computer Science and Information Technology, West Pomeranian University of Technology, Szczecin,  al.~Piastów~17, 70-310 Szczecin, Poland (e-mail: atariov@zut.edu.pl).}}

\markboth{Journal of \LaTeX\ Class Files, Vol. 14, No. 8, August 2015}
{Shell \MakeLowercase{\textit{et al.}}: Bare Demo of IEEEtran.cls for IEEE Journals}
\maketitle

\begin{abstract}
Fast Fourier transform algorithms are an arsenal of effective tools for solving various problems of analysis and high-speed processing of signals of various natures. Almost all of these algorithms are designed to process sequences of complex-valued data when each element of the sequence represents a single whole. However, in some cases, it is more advantageous to represent each element of the input and output sequences by a pair of real numbers. Such a need arises, for example, when further post-processing of spectral coefficients is carried out through two independent channels. Taking into account the noted need, the article proposes an algorithm for fast complex-valued discrete Fourier transform with separate real and imaginary inputs/outputs. A vector-matrix computational procedure is given that allows one to adequately describe and formalize the sequence of calculations when implementing the proposed algorithm.
\end{abstract}


\IEEEpeerreviewmaketitle

\section{Introduction}

\IEEEPARstart{T}{he} Fast Fourier Transform (FFT) is a fundamental family of algorithms in digital information processing, the best-known version of which was published by J. Cooley and  J. Tukey~\cite{bib01}. Indeed, the Cooley-Tukey algorithm is perhaps the most widely used today for analyzing and processing digital or discrete data. The original FFT algorithm processed sequences of signal samples whose length was equal to a power of two. However, the diversity of the problems being solved led to the development of new algorithms for calculating the discrete Fourier transform, which made it possible to eliminate this limitation and significantly expanded the family of fast algorithms.  With the development of fast computing, prime factor FFT algorithms~\cite{bib02, bib03, bib04} (also called the Good–Thomas algorithms), split-radix FFT algorithms~\cite{bib05, bib06}, vector-radix FFT algorithms~\cite{bib07, bib08}, split-vector radix algorithms~\cite{bib09}, and mixed-radix FFT algorithms~\cite{bib10} have emerged. 

Modifications of algorithms have appeared, named after their authors, such as the Rader-Brenner FFT~\cite{bib11} Stockham FFT algorithm~\cite{bib12}, Bluestein's FFT algorithm (also known as the chirp-z-transform algorithm)~\cite{bib13}, Pease FFT algorithm~\cite{bib14}, Winograd Fourier transform algorithms~\cite{bib15}, Brunn’s FFT algorithm~\cite{bib16, bib17}, etc. As a rule, all the listed methods consider each complex-valued sample as a single whole. Thus, in traditional FFT algorithms, the input data is represented by a vector, each element of which is a complex number. However, in some cases, it becomes necessary to represent the input and output data in such a way that the real and imaginary parts of the complex numbers are fed to the inputs of the algorithm, processed, and output separately. This paper is devoted to the development of such an algorithm.

\section{Preliminaries}

Mathematically, the $k$-th spectral component of the $N$-point Discrete Fourier transform (DFT) can be calculated as follows: 
\begin{equation}
y_k = \sum_{n=0}^{N-1} x_n \exp \left ( -j\frac{2\pi n k}{N} \right ),\quad k=0, 1, \dots , N-1
\label{eq-1}
\end{equation}
where $\{x_n\}\in \mathbb{C} , n=0, \dots, N-1$, is the sequence of complex-valued input data; $N\in \mathbb{R}$  is the number of signal samples;
\begin{gather*}
x_n=x_n^{(r)}+j x_n^{(i)};\quad y_n=x_y^{(r)}+j x_y^{(i)};
\end{gather*}
where $j=\sqrt{-1}$ and $x_n^{(r)}, x_n^{(i)}, y_n^{(r)}, y_n^{(i)} \in \mathbb{R}.$

In matrix-vector notation the DFT can be represented as follows:
\begin{equation}
  {\bf{y}}_{N} = {\bf{E}}_{N}\, {\bf{y}}_{N},
\label{eq2}
\end{equation}
where
\begin{equation*}
{\bf{E}}_N=
\begin{bmatrix}
               w^{0,0} & w^{0,1} & \cdots & w^{0,(N-1)}\\
               w^{1,0} & w^{1,1} & \cdots & w^{1,(N-1)}\\
               \vdots    &    \vdots  & \ddots & \vdots \\ 
               w^{(N-1),0} & w^{(N-1),1} & \cdots & w^{(N-1),(N-1)}
\end{bmatrix},
\end{equation*}
where
\begin{equation*}
w^{k,n} = \exp \left ( -j\frac{2\pi n k}{N} \right ),
\end{equation*}\vspace{-1.5em}
\begin{gather*}
{\bf{x}}_N = {\left [ x_0, x_1, \dots, x_{N-1} \right ]}^{T},\\ {\bf{y}}_N = {\left [ y_0, y_1, \dots, y_{N-1} \right ]}^{T}.
\end{gather*}

The idea behind all fast discrete Fourier transform algorithms is to use the remarkable properties of the matrix ${\bf{E}}_N$   to find a factorization that reduces the number of arithmetic operations required. It should be noted that there are different approaches to factorization of the matrix ${\bf{E}}$, which are the basis for various modifications of the FFT algorithms.

It is easy to see that expression~\eqref{eq2} can be rewritten as follows:
\begin{equation}
  {\bf{y}}_{N}^{(r)}+j{\bf{y}}_{N}^{(i)} = \left( {\bf{C}}_{N} - j {\bf{S}}_{N}\right) \left( {\bf{x}}_{N}^{(r)}+j{\bf{x}}_{N}^{(i)} \right)
\label{eq3}
\end{equation}
where
\begin{equation*}
{\bf{C}}_N=
\begin{bmatrix}
               c^{0,0} & c^{0,1} & \cdots & c^{0,(N-1)}\\
               c^{1,0} & c^{1,1} & \cdots & c^{1,(N-1)}\\
               \vdots    &    \vdots  & \ddots & \vdots \\ 
               c^{(N-1),0} & c^{(N-1),1} & \cdots & c^{(N-1),(N-1)},
\end{bmatrix},
\end{equation*}
\begin{equation*}
{\bf{S}}_N=
\begin{bmatrix}
               s^{0,0} & s^{0,1} & \cdots & s^{0,(N-1)}\\
               s^{1,0} & s^{1,1} & \cdots & s^{1,(N-1)}\\
               \vdots    &    \vdots  & \ddots & \vdots \\ 
               s^{(N-1),0} & s^{(N-1),1} & \cdots & s^{(N-1),(N-1)},
\end{bmatrix},
\end{equation*}
\begin{equation*}
c_{k,n}=\cos\frac{2\pi k n}{N}, \quad s_{k,n}=\sin\frac{2\pi k n}{N},
\end{equation*}
and
\begin{gather*}
{\bf{x}}_N^{(r)} = {\left [ x_0^{(r)}, x_1^{(r)}, \dots, x_{N-1}^{(r)} \right ]}^{T},\\ 
{\bf{x}}_N^{(i)} = {\left [ x_0^{(i)}, x_1^{(i)}, \dots, x_{N-1}^{(i)} \right ]}^{T},\\
{\bf{y}}_N^{(r)}= {\left [ y_0^{(r)}, y_1^{(r)}, \dots, y_{N-1}^{(r)} \right ]}^{T},\\  
{\bf{y}}_N^{(i)} = {\left [ y_0^{(i)}, y_1^{(i)}, \dots, y_{N-1}^{(i)} \right ]}^{T}.
\end{gather*}

It is easy to see that the calculations corresponding to expression~\eqref{eq2} can be performed as follows:
\begin{equation}
\begin{bmatrix} {\bf{y}}_N^{(r)}\\ {\bf{y}}_N^{(i)} \end{bmatrix}
= {\bf{E}}_{2N}
\begin{bmatrix} {\bf{x}}_N^{(r)}\\ {\bf{x}}_N^{(i)} \end{bmatrix}
\end{equation}
where
\begin{equation}
{\bf{E}}_{2N} = 
\begin{bmatrix} {\bf{C}}_{N} & {\bf{S}}_{N}\\ -{\bf{S}}_{N} & {\bf{C}}_{N} \end{bmatrix}
\end{equation}

The matrix ${\bf{E}}_{2N}$ has a benchmark block structure that allows finding different variants of factorization, reducing the number of arithmetic operations required to calculate the matrix-vector product. Below we will show one of such possibilities, which allows synthesizing an algorithm with independent real and imaginary inputs/outputs

\section{The algorithm}

We will first establish that in all subsequent discussions, for the sake of certainty, we will assume that the number of complex-valued elements of the original data sequence is equal to a power of two, i.e. $N=2^m$, $m \geq 2$ is an integer. 

We form two vectors of double length containing pairwise grouped real and imaginary parts of the elements of the input and output complex-valued sequences, respectively:
\begin{gather*}
{\bf{x}}_{2N}^{(r)}{\!} ={\!} {\left [ x_0^{(r)}{\!}, x_0^{(i)}{\!}, x_1^{(r)}{\!}, x_1^{(i)}{\!}, \dots, x_{N-1}^{(r)}, x_{N-1}^{(i)} \right ]}^{T}{\!},  \\
{\bf{y}}_{2N}^{(r)}{\!} ={\!} {\left [ y_0^{(r)}{\!}, y_0^{(i)}{\!}, y_1^{(r)}{\!}, y_1^{(i)}{\!}, \dots, y_{N-1}^{(r)}, y_{N-1}^{(i)} \right ]}^{T}{\!}.
\end{gather*}

Additionally, we introduce into consideration the matrix of a pairwise bit-reversal permutation
\begin{equation*}
{\bf S}_{2^{m+1}} = {\bf S}_{2^{m}} \otimes {\bf I}_2,
\end{equation*}
where ${\bf I}_N$ is identity matrix of order $N$, $\otimes$ is the Kronecker product sign~\cite{bib18}, and ${\bf S}_{2^{m}}$ is the bit-reversal permutation matrix with entries
\begin{equation*}
s_{k+1, {\langle n \rangle}_2}= 
\begin{cases} 1, & \text{if}\quad  k=n \in \overline{0, N-1}\\ 0, & \text{if}\quad k\neq n \end{cases}
\end{equation*}
and ${\langle n \rangle}_2$ is a bit-reversal representation of number $n$ performed according to the rule:
\begin{gather*}
     \text{if}\quad  n = \sum_{i=0}^{m-1} n_i 2^i,\quad
        \text{then}\quad  {\langle n\rangle}_2 = \sum_{i=0}^{m-1} n_{m-1-i} 2^i,\\[2pt]
       m=\log{_2}N, \quad n_i \in 0, 1.
\end{gather*}

Simply put, matrix ${\bf S}_{2^{m}}$ contains exactly one unit in each row, in the column whose number corresponds to the bit-reversal representation of the row number.

Considering the principles of formation of input vectors and omitting cumbersome derivations, the final complex-valued FFT algorithm with separate real and imaginary inputs and outputs can be represented using the following matrix-vector computational procedure:
\begin{equation}
{\bf y}_{2N} = {\bf W}^{(m)}_{2^{m+1}} {\bf D}^{(m)}_{2^{m+1}} \cdots {\bf W}^{(1)}_{2^{m+1}} 
                      {\bf D}^{(1)}_{2^{m+1}}  {\bf S}_{2^{m+1}}\, {\bf x}_{2N},\vspace{4pt}
\label{eq6}
\end{equation}
where
\begin{gather*}
{\bf W}^{(i)}_{2^{m+1}} = {\bf I}_{2^{m-1}} \otimes {\bf H}_2 \otimes {\bf I}_{2^{i}}, \\
          {\bf D}^{(i)}_{2^{m+1}}={\bf I}_{2^{m-1}} \otimes {\bf D}_{2^{i+1}},\\[6pt]
{\bf D}_{2^{i+1}} = \mbox{diag} \left( {\bf I}_{2R}, {\boldsymbol \omega}^{(i, 0)}_{2}, \dots, {\boldsymbol \omega}^{(i, R-1)}_{2} \right),
\quad R=2^{i-1},
\end{gather*}
\begin{equation*}
{\bf H}_{2}=\begin{bmatrix}1&1\\1&-1\end{bmatrix}, \quad 
{\boldsymbol \omega}^{(i, q)}_{2}=\begin{bmatrix} \cos\dfrac{2\pi q}{2^i} & \sin\dfrac{2\pi q}{2^i}\\[8pt]
                                                                 -\sin\dfrac{2\pi q}{2^i} & \cos\dfrac{2\pi q}{2^i}\end{bmatrix}.   
\end{equation*}

Let us consider an example. Let $N=2^3$. In this case, the matrix-vector computational procedure describing the proposed algorithm will be written as follows:
\begin{equation}
{\bf y}_{16} = {\bf W}^{(3)}_{2^{4}} {\bf D}^{(3)}_{2^{4}} \cdots {\bf W}^{(1)}_{2^{4}} 
                      {\bf D}^{(1)}_{2^{4}}  {\bf S}_{2^{4}}\, {\bf x}_{16},\vspace{-4pt}
\label{eq7}
\end{equation}
where\vspace{-8pt}
\begin{gather*}
{\bf{x}}_{16}^{(r)}{\!} ={\!} {\left [ x_0^{(r)}{\!}, x_0^{(i)}{\!}, x_1^{(r)}{\!}, x_1^{(i)}{\!}, \dots, x_{7}^{(r)}, x_{7}^{(i)} \right ]}^{T}{\!},  \\
{\bf{y}}_{16}^{(r)}{\!} ={\!} {\left [ y_0^{(r)}{\!}, y_0^{(i)}{\!}, y_1^{(r)}{\!}, y_1^{(i)}{\!}, \dots, y_{7}^{(r)}, y_{7}^{(i)} \right ]}^{T}{\!},\\[-18pt]
\end{gather*}
\begin{gather*}
{\bf S}_{2^4}={\bf S}_{2^3} \otimes {\bf I}_{2}=\mbox{\hspace{135pt}}\\
=\begin{bmatrix}
     1&&&&&&& \\ 
     &&&&1&&& \\ 
     &&1&&&&& \\
     &&&&&&1& \\
     &1&&&&&& \\
     &&&&&1&& \\ 
    &&&1&&&& \\ 
    &&&&&&&1
\end{bmatrix}
\otimes \begin{bmatrix} 1&\\&1\end{bmatrix}.
\end{gather*}

Now we will show how to obtain the remaining matrices. Let $i=1$, $R=2^{1-1}=1$. Then
\begin{gather*}
          {\bf D}^{(1)}_{2^{3+1}}={\bf I}_{2^{3-1}} \otimes {\bf D}_{2^{1+1}}={\bf I}_{2^{2}} \otimes {\bf D}_{2^{2}},\\
{\bf D}_{2^{2}} = \mbox{diag} \left( {\bf I}_{2}, {\boldsymbol \omega}^{(1, 0)}_{2}\right),\\
 {\bf D}^{(1)}_{2^{4}}={\bf I}_{4} \otimes  \mbox{diag} \left( {\bf I}_{2}, {\boldsymbol \omega}^{(1, 0)}_{2}\right),
\quad{\boldsymbol \omega}^{(1, 0)}_{2}=\begin{bmatrix}1&\\&1\end{bmatrix}
\end{gather*}
So, the matrix ${\bf D}^{(1)}_{2^{4}}$ is
\begin{equation*}
{\bf D}^{(1)}_{2^{4}}=\mbox{diag}(1, 1, 1, 1, 1, 1, 1, 1, 1, 1, 1, 1, 1, 1, 1, 1),
\end{equation*}
whereas
\begin{gather*}
{\bf W}^{(1)}_{2^4}={\bf I}_{2^{3-1}} \otimes {\bf H}_{2} \otimes {\bf I}_{2^{1}}=
                                 {\bf I}_{4} \otimes {\bf H}_{2} \otimes {\bf I}_{2}=\mbox{\hspace{45pt}}\\[2pt]
=\begin{bmatrix}
    1&&&\\    &1&&\\    &&1&\\    &&&1
\end{bmatrix}
\otimes
\begin{bmatrix}
    1&1\\    1&-1
\end{bmatrix}
\otimes \begin{bmatrix} 1&\\&1\end{bmatrix}.
\end{gather*}

Let now $i=2$, $R=2^{2-1}=2$. Then
\begin{gather*}
          {\bf D}^{(2)}_{2^{3+1}}={\bf I}_{2^{3-2}} \otimes {\bf D}_{2^{2+1}}={\bf I}_{2} \otimes {\bf D}_{2^{3}},\\[2pt]
{\bf D}_{2^{3}} = \mbox{diag} \left( {\bf I}_{4}, {\boldsymbol\omega}^{(2, 0)}_{2}, {\boldsymbol \omega}^{(2, 1)}_{2}\right),\\[3pt]
{\boldsymbol \omega}^{(2, 0)}_{2}=\begin{bmatrix}1&\\&1\end{bmatrix},\quad 
{\boldsymbol \omega}^{(2, 1)}_{2}=\begin{bmatrix}&1\\-1&\end{bmatrix}.
\end{gather*}
\begin{equation*}
 {\bf D}^{(2)}_{2^{4}}={\bf I}_{2} \otimes  \mbox{diag} \left( {\bf I}_{4}, {\boldsymbol \omega}^{(2, 0)}_{2},{\boldsymbol \omega}^{(2, 1)}_{2}\right)= \mbox{\hspace{75pt}}
\end{equation*}
\begin{equation*}
=\begin{bmatrix}1&\\&1\end{bmatrix} \otimes
\begin{bmatrix}
{\begin{bmatrix} 1&&&\\   &1&&\\   &&1&\\   &&&1\\ \end{bmatrix}}
 &
{\begin{matrix}   &&&\\   &&&\\   &&&\\   &&&\\ \end{matrix}} \\
{\begin{matrix} &&& \\ &&&\\ \end{matrix}}
 &
\hspace{-20pt}{\begin{bmatrix} 1&\\ &1\\ \end{bmatrix}}
&
{\begin{matrix} &\\ &\\ \end{matrix}}\\
{\begin{matrix} &&& \\ &&&\\ \end{matrix}}
 &
{\begin{matrix} &\\ &\\ \end{matrix}}
&
\hspace{-20pt}{\begin{bmatrix} &1\\ -1&\\ \end{bmatrix}}\\
\end{bmatrix}
\end{equation*}


\begin{gather*}
{\bf W}^{(2)}_{2^{4}}={\bf I}_{2^{3-2}} \otimes {\bf H}_{2} \otimes {\bf I}_{2^{2}}=
                                 {\bf I}_{2} \otimes {\bf H}_{2} \otimes {\bf I}_{4}=\mbox{\hspace{33pt}}\\
=\begin{bmatrix} 1&\\&1\end{bmatrix}
\otimes
\begin{bmatrix}
    1&1\\    1&-1
\end{bmatrix}
\otimes \begin{bmatrix}
    1&&&\\    &1&&\\    &&1&\\    &&&1
\end{bmatrix}.
\end{gather*}

And finally, let now $i=3$, $R=2^{3-1}=4$. Then
\begin{gather*}
    {\bf D}^{(2)}_{2^{3+1}}={\bf I}_{2^{3-3}} \otimes {\bf D}_{2^{3+1}}={1} \otimes {\bf D}_{2^{4}}={\bf D}_{2^{4}},\\[2pt]
{\bf D}_{2^{4}} = \mbox{diag} \left( {\bf I}_{8}, {\boldsymbol \omega}^{(3, 0)}_{2}, 
{\boldsymbol \omega}^{(3, 1)}_{2}, {\boldsymbol \omega}^{(3, 2)}_{2}, {\boldsymbol \omega}^{(3, 3)}_{2}\right),\\
\end{gather*}
\begin{gather*}
{\boldsymbol \omega}^{(3, 0)}_{2}=\begin{bmatrix}1&\\&1\end{bmatrix},\quad 
{\boldsymbol \omega}^{(3, 2)}_{2}=\begin{bmatrix}&1\\-1&\end{bmatrix}\\[2pt]
{\boldsymbol \omega}^{(3, 1)}_{2}=\begin{bmatrix}1/\sqrt{2}&1/\sqrt{2}\\[2pt]-1/\sqrt{2}&1/\sqrt{2}\end{bmatrix},\quad 
{\boldsymbol \omega}^{(3, 3)}_{2}=\begin{bmatrix}1/\sqrt{2}&1/\sqrt{2}\\[2pt]-1/\sqrt{2}&1/\sqrt{2}\end{bmatrix},\\[4pt]
{\bf W}^{(3)}_{2^{4}}={\bf I}_{2^{3-3}} \otimes {\bf H}_{2} \otimes {\bf I}_{2^{3}}=
                                    {\bf H}_{2} \otimes {\bf I}_{8}=\mbox{\hspace{53pt}}\\
=\begin{bmatrix} 1&1\\1&-1\end{bmatrix}
\otimes \begin{bmatrix}
    1&&&&&&&\\    &1&&&&&&\\    &&1&&&&&\\    &&&1&&&&\\ 
    &&&&1&&&\\    &&&&&1&&\\    &&&&&&1&\\    &&&&&&&1\\
\end{bmatrix},\\[-20pt]
\end{gather*}
\begin{gather*}
{\bf D}_{2^{4}}=
   \begin{bmatrix}
   1&&&&&&&&&\\   &1&&&&&&&&\\   &&1&&&&&&&\\   &&&1&&&&&&\\   &&&&1&&&&&\\
   &&&&&1&&&&\\   &&&&&&1&&&\\   &&&&&&&1&&\\   &&&&&&&&1&\\   &&&&&&&&&1\\
   \end{bmatrix}
\oplus \mbox{\hspace{30pt}}
\end{gather*}
\begin{equation}
\oplus
\begin{bmatrix}
{\begin{bmatrix}1/\sqrt{2}&1/\sqrt{2}\\[2pt]-1/\sqrt{2}&1/\sqrt{2}\end{bmatrix}}
 &
{\begin{matrix}   &&&\\   &&&\\ \end{matrix}} \\
{\begin{matrix} & \\ &\\ \end{matrix}}
 &
\hspace{-10pt}{\begin{bmatrix} &1\\ -1&\\ \end{bmatrix}}
& {\begin{matrix} &\\ &\\ \end{matrix}}\\
{\begin{matrix} & \\ & \end{matrix}} & {\begin{matrix} & \\ & \end{matrix}} &
\hspace{-10pt}{\begin{bmatrix}1/\sqrt{2}&1/\sqrt{2}\\[2pt]-1/\sqrt{2}&1/\sqrt{2} \end{bmatrix}}\\
\end{bmatrix}
\label{eq8}
\end{equation}
Note that the symbol $\oplus$ used in~\eqref{eq8} denotes the direct sum of the two matrices\cite{bib17}.

\begin{figure}[tb!]
\centerline{\includegraphics[width=\columnwidth]{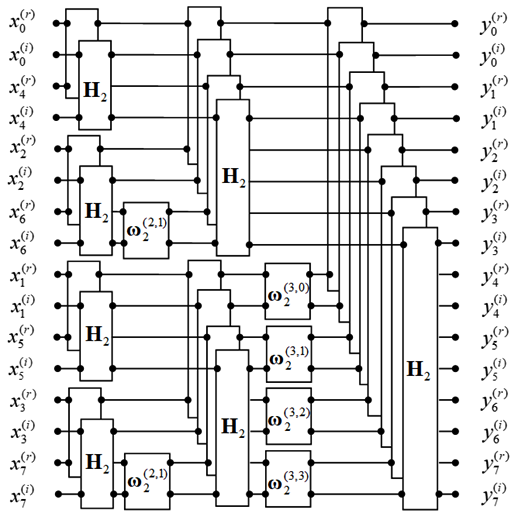}}
\caption{Data flow diagram of the algorithm for 8-point decimation in time FFT with separate real and imaginary inputs/outputs.}\label{Fig_1}
\end{figure}

Fig.~\ref{Fig_1} shows the data flow diagram corresponding to the example considered. Here, the straight lines represent data transfer operations, and the rectangles represent the multiplication of a two-element input vector by a matrix whose symbol is inscribed inside the rectangle. The data flow is from left to right. 

The considered algorithm is a modification of the Cooley-Tukey algorithm with decimation in time. In turn, the common expression for a similar algorithm with decimation in frequency is written as follows:
\begin{equation}
{\bf y}_{2N}={\bf S}_{2^{m+1}}{\bf D}^{(1)}_{2^{m+1}}{\bf W}^{(1)}_{2^{m+1}}\dots 
                    {\bf D}^{(m)}_{2^{m+1}}{\bf W}^{(m)}_{2^{m+1}}{\bf x}_{2N}
\label{eq9}
\end{equation}
and the 8-point transform will take the following form:
\begin{equation}
{\bf y}_{16}={\bf S}_{2^{4}}{\bf D}^{(1)}_{2^{4}}{\bf W}^{(1)}_{2^{4}}\dots 
                    {\bf D}^{(3)}_{2^{4}}{\bf W}^{(3)}_{2^{4}}{\bf x}_{16}.
\label{eq10}
\end{equation}

\begin{figure}[tb!]
\centerline{\includegraphics[width=\columnwidth]{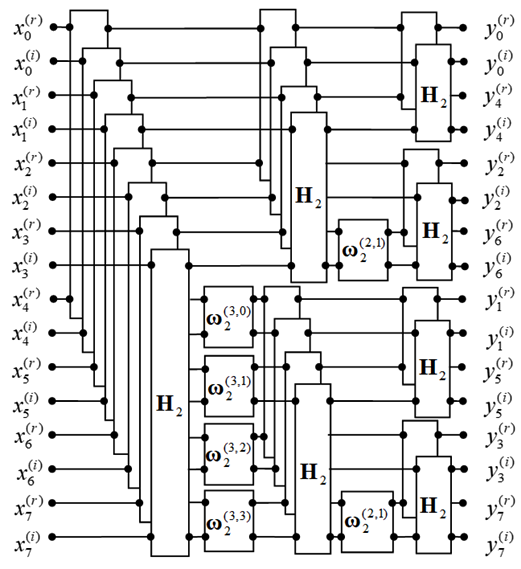}}
\caption{Data flow diagram of the algorithm for 8-point decimation in frequency FFT with separate real and imaginary inputs/outputs.}\label{Fig_2}
\end{figure}

Fig.~\ref{Fig_2} shows the data flow diagram corresponding to the 8-point decimation in frequency FFT algorithm with separate real and imaginary inputs/outputs. It can be seen that the data flow diagram in Fig.~\ref{Fig_2} is a mirror image of the diagram shown in Fig.~\ref{Fig_1}. If you don't read it carefully, it might seem like this is an inverse FFT diagram, but this algorithm has nothing in common with the inverse FFT; it is also a forward FFT.

\section{Conclusion}

The paper presents fast algorithms for complex-valued discrete Fourier transform with separate real and imaginary inputs and outputs. By and large, they do not represent anything fundamentally new compared to traditional Cooley-Tukey algorithms, the idea is the same. However, they differ from traditional Cooley-Tukey algorithms in that the real and imaginary parts of complex-valued signal samples are processed separately. We simply move from the domain of complex operations to the domain of real operations, which simplifies the organization of calculations. The computational complexity of the presented algorithms is the same as that of the Cooley-Tukey algorithms, since these are the same algorithms. Their main difference and advantage over classical FFT algorithms is the ability to unify the description of the implementation of the computational process in the case when complex-valued inputs and outputs are represented and processed separately.

\bibliographystyle{IEEEtran}

\end{document}